\documentclass[12pt,a4paper]{article}
\usepackage{epsfig}                    
\usepackage{amsmath}
\begin{document}
\textwidth=135mm
\textheight=200mm
\tolerance=100000

\begin{center}
{\bfseries Solar neutrino with Borexino: results and perspectives
\footnote{{\small Talk at the International Workshop on Prospects of Particle Physics, "Neutrino Physics and Astrophysics", Valday, Russia, January 26 - February 02, 2014.}}}
\vskip 5mm
O.~Smirnov$^{\mbox{b}}$, 
G.~Bellini$^{\mbox{a}}$,
J.~Benziger$^{\mbox{k}}$, 
D.~Bick$^{\mbox{s}}$, 
G.~Bonfini$^{\mbox{e}}$,
D.~Bravo$^{\mbox{q}}$,
B.~Caccianiga$^{\mbox{a}}$,
F.~Calaprice$^{\mbox{l}}$, 
A.~Caminata$^{\mbox{c}}$, 
P.~Cavalcante$^{\mbox{e}}$, 
A.~Chavarria$^{\mbox{l}}$, 
A.~Chepurnov$^{\mbox{r}}$,
D.~D{\textquoteright}Angelo$^{\mbox{a}}$,
S.~Davini$^{\mbox{t}}$, 
A.~Derbin$^{\mbox{m}}$,
A.~Empl$^{\mbox{t}}$,
A.~Etenko$^{\mbox{g}}$,
K.~Fomenko$^{\mbox{{b,e}}}$, 
D.~Franco$^{\mbox{{h}}}$,
G.~Fiorentini$^{\mbox{{v}}}$,
C.~Galbiati$^{\mbox{l}}$,
S.~Gazzana$^{\mbox{e}}$, 
C.~Ghiano$^{\mbox{c}}$, 
M.~Giammarchi$^{\mbox{a}}$, 
M.~G\"{o}ger-Neff$^{\mbox{n}}$, 
A.~Goretti$^{\mbox{l}}$,
C.~Hagner$^{\mbox{s}}$, 
E.~Hungerford$^{\mbox{t}}$,
Aldo Ianni$^{\mbox{e}}$,
Andrea Ianni$^{\mbox{l}}$,
V.~Kobychev$^{\mbox{f}}$, 
D.~Korablev$^{\mbox{b}}$, 
G.~Korga$^{\mbox{t}}$, 
D.~Kryn$^{\mbox{h}}$,
M.~Laubenstein$^{\mbox{e}}$, 
B.~Lehnert$^{\mbox{w}}$,
T.~Lewke$^{\mbox{n}}$,
E.~Litvinovich$^{\mbox{g}}$, 
F.~Lombardi$^{\mbox{e}}$, 
P.~Lombardi$^{\mbox{a}}$, 
L.~Ludhova$^{\mbox{a}}$, 
G.~Lukyanchenko$^{\mbox{g}}$,
I.~Machulin$^{\mbox{g}}$,
S.~Manecki$^{\mbox{q}}$, 
W.~Maneschg$^{\mbox{i}}$, 
F.~Mantovani$^{\mbox{{v}}}$,
S.~Marcocci$^{\mbox{c}}$, 
Q.~Meindl$^{\mbox{n}}$, 
E.~Meroni$^{\mbox{a}}$, 
M.~Meyer$^{\mbox{s}}$, 
L.~Miramonti$^{\mbox{a}}$,
M.~Misiaszek$^{\mbox{d}}$, 
P.~Mosteiro$^{\mbox{l}}$, 
V.~Muratova$^{\mbox{m}}$,
L.~Oberauer$^{\mbox{n}}$,
M.~Obolensky$^{\mbox{h}}$, 
F.~Ortica$^{\mbox{j}}$, 
K.~Otis$^{\mbox{p}}$,
M.~Pallavicini$^{\mbox{c}}$, 
L.~Papp$^{\mbox{{q}}}$,
L.~Perasso$^{\mbox{c}}$, 
A.~Pocar$^{\mbox{p}}$,
G.~Ranucci$^{\mbox{a}}$, 
A.~Razeto$^{\mbox{e}}$, 
A.~Re$^{\mbox{a}}$, 
B.~Ricci$^{\mbox{{v}}}$,
A.~Romani$^{\mbox{j}}$, 
N.~Rossi$^{\mbox{e}}$,
R.~Saldanha$^{\mbox{l}}$, 
C.~Salvo$^{\mbox{c}}$, 
S.~Sch\"onert$^{\mbox{{n}}}$, 
H.~Simgen$^{\mbox{i}}$, 
M.~Skorokhvatov$^{\mbox{g}}$, 
A.~Sotnikov$^{\mbox{b}}$, 
S.~Sukhotin$^{\mbox{g}}$, 
Y.~Suvorov$^{\mbox{{u,g}}}$,
R.~Tartaglia$^{\mbox{e}}$,
G.~Testera$^{\mbox{c}}$,
D.~Vignaud$^{\mbox{h}}$, 
R.B.~Vogelaar$^{\mbox{q}}$, 
F.~von Feilitzsch$^{\mbox{n}}$,
H.~Wang$^{\mbox{u}}$, 
J.~Winter$^{\mbox{n}}$, 
M.~Wojcik$^{\mbox{d}}$, 
A.~Wright$^{\mbox{l}}$, 
M.~Wurm$^{\mbox{s}}$,
O.~Zaimidoroga$^{\mbox{b}}$, 
S.~Zavatarelli$^{\mbox{c}}$, 
K.~Zuber$^{\mbox{w}}$, 
and G.~Zuzel$^{\mbox{d}}$.

(Borexino Collaboration)

\vskip 5mm
{\small {\it 
$^{\mbox{a}}$                         
Dipartimento di Fisica, Universit\`{a} degli Studi e INFN, Milano 20133, Italy\\
$^{\mbox{b}}$                         
Joint Institute for Nuclear Research, Dubna 141980, Russia\\
$^{\mbox{c}}$                         
Dipartimento di Fisica, Universit\`{a} e INFN, Genova 16146, Italy\\
$^{\mbox{d}}$                         
M. Smoluchowski Institute of Physics, Jagellonian University, Crakow, 30059, Poland\\
$^{\mbox{e}}$                         
INFN Laboratori Nazionali del Gran Sasso, Assergi 67010, Italy\\
$^{\mbox{f}}$                         
Kiev Institute for Nuclear Research, Kiev 06380, Ukraine\\
$^{\mbox{g}}$                         
NRC Kurchatov Institute, Moscow 123182, Russia\\
$^{\mbox{h}}$                         
APC, Univ. Paris Diderot, CNRS/IN2P3, CEA/Irfu, Obs. de Paris, Sorbonne Paris Cit\'e, France\\
$^{\mbox{i}}$                         
Max-Plank-Institut f\"{u}r Kernphysik, Heidelberg 69029, Germany\\
$^{\mbox{j}}$                         
Dipartimento di Chimica, Universit\`{a} e INFN, Perugia 06123, Italy\\
$^{\mbox{k}}$                         
Chemical Engineering Department, Princeton University, Princeton, NJ 08544, USA\\
$^{\mbox{l}}$                         
Physics Department, Princeton University, Princeton, NJ 08544, USA\\
$^{\mbox{m}}$                         
St. Petersburg Nuclear Physics Institute, Gatchina 188350, Russia\\
$^{\mbox{n}}$                         
Physik Department, Technische Universit\"{a}t M\"{u}nchen, Garching 85747, Germany\\
$^{\mbox{p}}$                         
Physics Department, University of Massachusetts, Amherst MA 01003, USA\\
$^{\mbox{q}}$                         
Physics Department, Virginia Polytechnic Institute and State University, Blacksburg, VA 24061, USA\\
$^{\mbox{r}}$                         
Lomonosov Moscow State University Skobeltsyn Institute of Nuclear Physics, Moscow 119234, Russia\\
$^{\mbox{s}}$                         
Institut f\"ur Experimentalphysik, Universit\"at Hamburg, Germany\\
$^{\mbox{t}}$                         
Department of Physics, University of Houston, Houston, TX 77204, USA\\
$^{\mbox{u}}$                         
Physics ans Astronomy Department, University of California Los Angeles (UCLA), Los Angeles, CA 90095, USA\\ 
$^{\mbox{v}}$                         
Dipartimento di Fisica e Scienze della Terra, Universit\`{a} degli Studi and INFN, Ferrara I-44122, Italy\\
$^{\mbox{w}}$                         
Institut f{\"u}r Kern- und Teilchenphysik, Technische Universit\"{a}t Dresden, Dresden 01069, Germany
}}
\end{center}
\vskip 5mm

\title{Solar neutrino with Borexino: results and perspectives.}

\begin{abstract}

Borexino is a unique detector able to perform measurement of solar
neutrinos fluxes in the energy region around 1 MeV or below due to
its low level of radioactive background. It was constructed at
the LNGS underground laboratory with a goal of solar $^{7}$Be neutrino
flux measurement with 5\% precision. The goal has been successfully
achieved marking the end of the first stage of the experiment. A number of other important
measurements of solar neutrino fluxes have been performed during the first stage.
Recently the collaboration conducted successful
liquid scintillator repurification campaign aiming to reduce main contaminants in the
sub-MeV energy range. With the new levels of radiopurity Borexino
can improve existing and challenge a number of new measurements including:
improvement of the results on the Solar and terrestrial neutrino fluxes
measurements; measurement of pp and CNO solar neutrino fluxes; search
for non-standard interactions of neutrino; study of the neutrino oscillations
on the short baseline with an artificial neutrino source (search for
sterile neutrino) in context of SOX project.
\end{abstract}




\section{Detector Borexino}

Borexino is a unique detector able to perform measurement of solar
neutrinos fluxes in the energy region around 1 MeV or below because
of its low level of radioactive background. After several years of
efforts and tests with the prototype CTF detector the design goals have been
reached and for some of the radioactive isotopes (internal $^{238}$U and $^{232}$Th)
largely exceeded. The low background is an essential condition to
perform the measurement: in fact solar neutrinos induced scintillations
cannot be distinguished on an event by event analysis from the ones
due to background. The energy shape of the solar neutrino is the main
signature that has to be recognized in the experimental energy spectrum
by a suitable fit procedure that includes the expected signal and
the background. The basic signature for the mono-energetic 0.862 MeV
$^{7}$Be neutrinos is the Compton-like edge of the recoil electrons
at 665 keV. 

The detector is located deep underground (approximately 3800 m of
water equivalent, mwe) in the Hall C of the Laboratori Nazionali del
Gran Sasso (Italy), where the muon flux is suppressed by a factor
of 10$^{6}$. The main goal of the experiment was the detection of
the monochromatic neutrinos that are emitted in the electron capture
of $^{7}$Be in the Sun with 5\% precision.

\begin{figure}
\begin{centering}
\includegraphics[width=0.55\paperwidth,height=0.4\paperwidth]{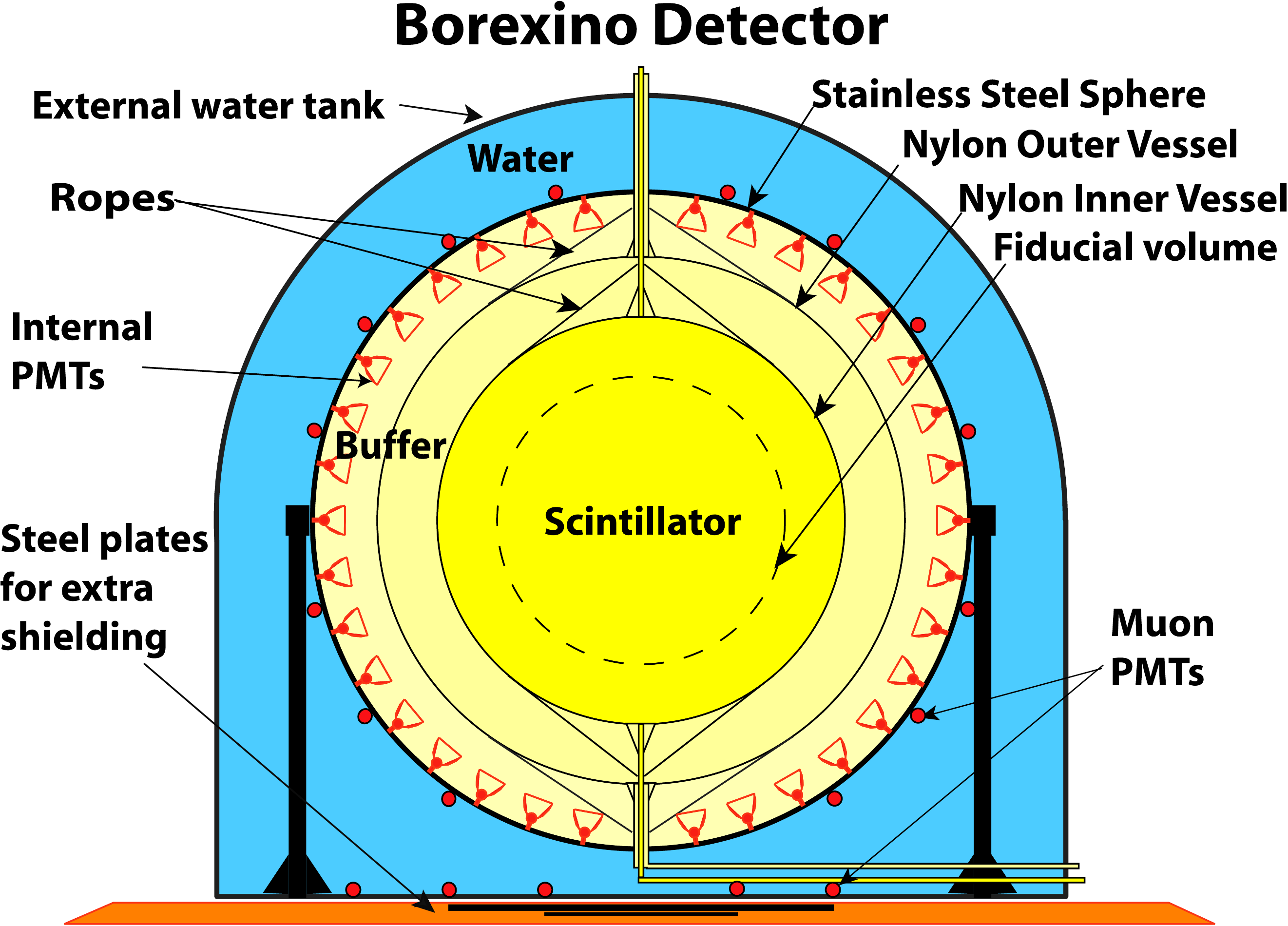}
\par\end{centering}
\caption{\label{Detector}Sketch of the Borexino detector. The base of the dome-like structure is 18 m in diameter.}
\end{figure}

The complete up to date technical description of the Borexino detector
has been reported in \cite{B02b} and \cite{Brx08}. The detector
is schematically depicted in Fig.\ref{Detector}. The inner part is
an unsegmented stainless steel sphere (SSS) that is both the container
of the scintillator and the mechanical support of the photomultipliers.
Within this sphere, two nylon vessels separate the scintillator volume
in three shells of radii 4.25 m, 5.50 m and 6.85 m, the latter being
the radius of the SSS itself. The inner nylon vessel (IV) contains
the liquid scintillator solution, namely PC (pseudocumene, 1,2,4-trimethylbenzene
C$_{6}$H$_{3}$(CH$_{3}$)$_{3}$) as a solvent and the fluor PPO
(2,5- diphenyloxazole, C$_{15}$H$_{11}$NO) as a solute at a concentration
of 1.5 g/l (0.17\% by weight). The second and the third shell contain
PC with a small amount (5 g/l) of DMP (dimethylphthalate) that is
added as a light quencher in order to further reduce the scintillation
yield of pure PC. The PC/PPO solution adopted as liquid scintillator
satisfies specific requirements: high scintillation yield ($\sim10^{4}$
photons/MeV), high light transparency (the mean free path is typically
8 m) and fast decay time ($\sim$3 ns), all essential for good energy
resolution, precise spatial reconstruction, and good discrimination
between $\beta$-like events and events due $\alpha$ particles. Furthermore,
several conventional petrochemical techniques are feasible to purify
the hundred of tons of fluids needed by Borexino.

The scintillation light is collected by 2212 photomutipliers (PMTs)
that are uniformly attached to the inner surface of the SSS. All but
384 photomultipliers are equipped with light concentrators that are
designed to reject photons not coming from the active scintillator
volume, thus reducing the background due to radioactive decays originating
in the buffer liquid or $\gamma$'s from the PMTs. The tank has a cylindrical base with
a diameter of 18 m and a hemispherical top with a maximum height of
16.9 m. The Water Tank (WT) is a powerful shielding against external
background ($\gamma$--rays and neutrons from the rock) and is also used as
a Cherenkov muon counter and muon tracker. The muon flux, although
reduced by a factor of $10^{6}$ by the 3800 m.w.e. depth of the Gran
Sasso Laboratory, is of the order of 1 m$^{-2}$ h$^{-1}$, corresponding
to about 4000 muons per day crossing the detector. This flux is well
above Borexino requirements and a strong additional reduction factor
(about $10^{4}$) is necessary. Therefore the WT is equipped with
208 photomultipliers that collect the Cherenkov light emitted by muons
in water. In order to maximize the light collection efficiency the
SSS and the interior of the WT surface are covered with a layer of
Tyvek, a white paper-like material made of polyethylene fibers.

The Borexino has an excellent energy resolution for its size, this
is the result of the high light yield of $\sim500$ p.e./MeV/2000
PMTs. The energy resolution (1$\sigma$) at the $^{7}$Be Compton edge energy (662
keV) is as low as 44 keV (or 6.6\%).

\section{Solar neutrino measurements at the first stage of the experiment}

Analysis of the Borexino data showed that the main goals concerning
the natural radioactivity have been achieved. The contamination of
the liquid scintillator with respect to the U/Th is at the level of
${10}^{-17}$ g/g; the contamination with $^{40}$K is at the level
of $10^{-19}\;$g/g ($10^{-15}$ in natural potassium); the $^{14}$C content is $2.7\pm0.1\times10^{-18}$
g/g with respect to the $^{12}$C. Among the other contamination sources
only $^{85}$Kr, $^{210}$Bi and $^{210}$Po have been identified.
The $^{85}$Kr counts $\sim$0.3 ev/day/tone, it is $\beta$- emitter
with 687 keV end-point. The $^{210}$Po is the most intense contamination
(with initial activity of 60 counts/day/tone), it decays emitting monoenergetic $\alpha$ with
5.3 MeV energy, the half-life time of the isotope is 134 days. The
residual contaminations do not obscure the expected neutrino signal,
the presence of the 862 keV monoenergetic $^{7}$Be solar neutrino
is clearly seen in the experimental spectrum. In such a way, the collaboration
succeeded to purify the liquid scintillator from residual natural
radioactive isotopes down to the levels much lower than was initially
envisaged for the $^{7}$Be neutrino measurement, which resulted in
broadening of the initial scientific scope of the experiment. 

\begin{figure}
\begin{centering}
\includegraphics[width=0.6\textwidth,height=0.5\textwidth]{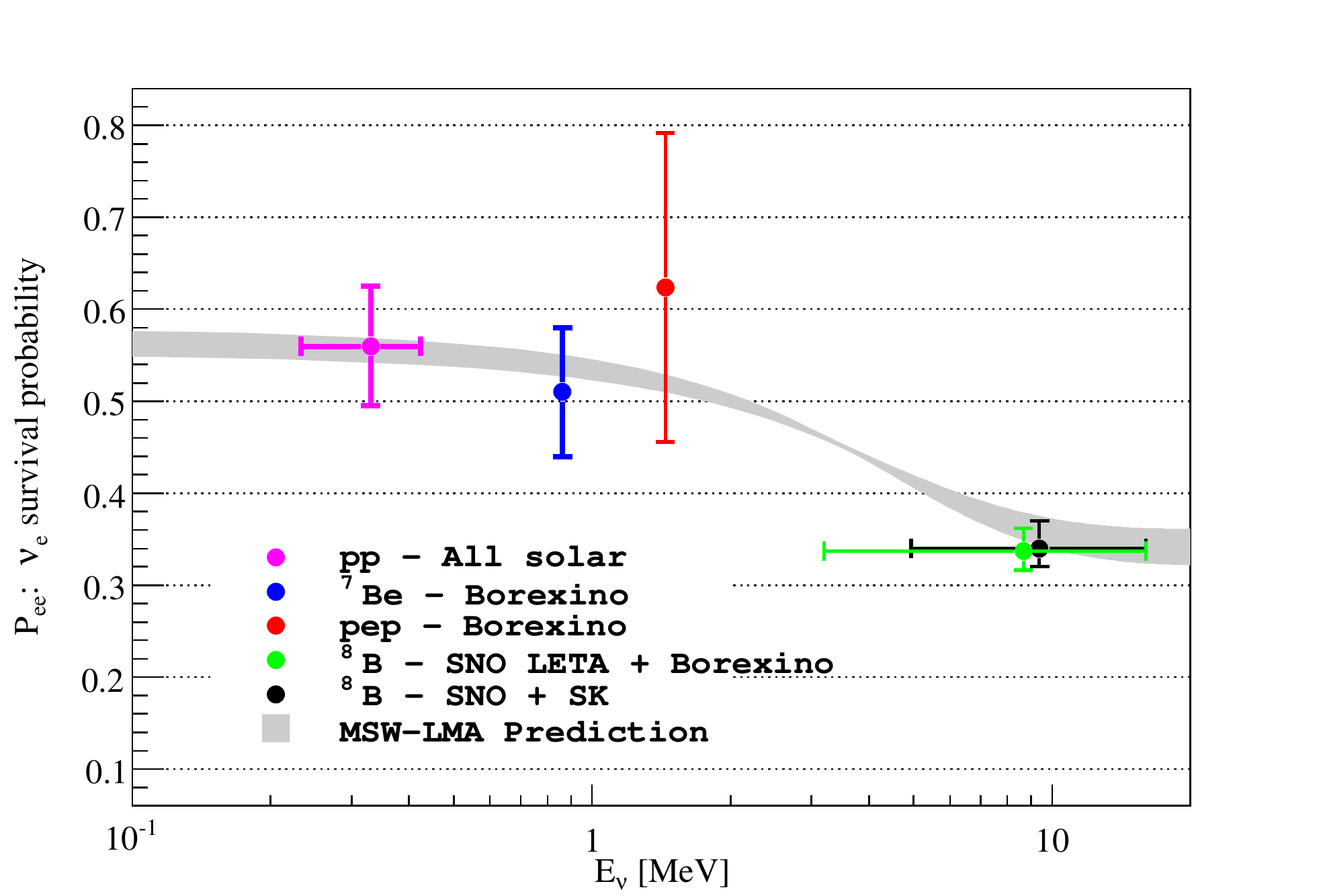}
\par\end{centering}

\caption{\label{Pee}electron neutrino survival probability, three points in
this plot are derived from the Borexino data. It should be noted that
the precision of the Be-7 neutrino flux itself is much better than
the Pee error at the corresponding point, this is related to the solar
model uncertainties. }

\end{figure}

The main goal of the experiment was the detection of the monochromatic
neutrinos that are emitted in the electron capture of $^{7}$Be in
the Sun with 5\% precision. This goal has been achieved during the
first stage of the experiment \cite{BOR07,Be708,Be7}. The Borexino
reported the first measurement of neutrino $^{8}$B neutrinos with
liquid scintillator detector with a 3 MeV threshold on electrons recoil
\cite{B8}. The stability of the detector allowed also to study the
day-night effect of the $^{7}$Be solar neutrino signal, thereby allowing
to completely exclude the LOW solution of the neutrino oscillation
based on solar data alone \cite{DN11}. Finally, the low background
of the detector, the refined analysis on threefold coincidences \cite{CNO06}
and the positronium discrimination method based on the positronium
formation study made it possible to explore the 1-2 MeV region with
unprecedented sensitivity. This led to the first observation of solar
neutrinos from the basic pep reaction \cite{PEP11}. In addition,
the best limit for the CNO production in a star has been established.
In this way, Borexino has completed direct detection of Be-7, pep
and B-8 solar neutrino components thereby providing complete evidence
of the transition from MSW and vacuum oscillation of the LMA solution
of the Solar Neutrino Problem (fig. \ref{Pee}).

\section{Solar neutrino program of the Borexino Phase II}

One of the goal of the Borexino experiment is the measurement of all
the solar neutrino fluxes, with the exception of the hep flux, too
faint for detection in Borexino. The $^{7}$Be, pep and $^{8}$B (this
last with the lowest threshold to date) have been already measured,
but the experimental uncertainties can be reduced. In addition Borexino
will try to measure the pp and CNO fluxes. 

\begin{table}

\begin{tabular}{|c|c|c|c|c|}
\hline 
$\nu$ flux & GS98 & AGS09 & cm$^{-2}$s$^{-1}$ & Experimental result\tabularnewline
\hline
\hline 
pep & $1.44\pm0.012$ & $1.47\pm0.012$ & $\times10^{8}$ & 1.6$\pm0.3$ Borexino\tabularnewline
\hline 
$^{7}$Be & $5.00\pm0.07$ & $4.56\pm0.07$ & $\times10^{9}$ & $4.87\pm0.24$ Borexino\tabularnewline
\hline 
$^{8}$B & $5.58\pm0.14$ & $4.59\pm0.14$ & $\times10^{6}$ & $5.2\pm0.3$ SNO+SK+Borexino+KamLAND\tabularnewline
 &  &  &  & $5.25\pm0.16_{-0.013}^{+0.011}$SNO-LETA\tabularnewline
\hline 
$^{13}$N & $2.96\pm0.14$ & $2.17\pm0.14$ & $\times10^{8}$ & \tabularnewline
\cline{1-4} 
$^{15}$O & $2.23\pm0.15$ & $1.56\pm0.15$ & $\times10^{8}$ & $<$7.4 Borexino (total CNO)\tabularnewline
\cline{1-4} 
$^{17}$F & $5.52\pm0.17$ & $3.40\pm0.16$ & $\times10^{8}$ & \tabularnewline
\hline
\end{tabular}

\caption{\label{SSMvsData}SSM predictions and current experimental results}

\end{table}

In table \ref{SSMvsData} the solar fluxes measured by Borexino so
far are compared with the SSM prediction, for low and high metallicity.
The experimental results agree, within the errors, with the SSM predictions,
but cannot distinguish between the two metallicities, due to the uncertainties
of the model and the experimental errors. It would be useful, at this
moment, to recall what the metallicity puzzle is. The solar surface
heavy element abundance has been calculated about ten years ago with
a 1D model, which uses data from spectroscopic observations of the
elements present in the photosphere (GS98 \cite{Grevesse}). This
model agrees with the helioseismology observations, namely the measurement
of the speed of the mechanical waves in the Sun. More recently a 3D
hydro-dynamical model (AGSS09 \cite{Asplund}) of the near-surface
solar convection, with improved energy transfer, has changed the Z/X
ratio with respect to the previous 1D treatment: 0.0178 (low metallicity)
to be compared with the previous 0.0229 (high metallicity). The 3D
model results perfectly reproduce the observed solar atmospheric line
(atomic and molecular) profiles and asymmetries, but are in clear
disagreement with the helioseismology data. At present there is no
satisfactory solution to this controversy \cite{Serenelli}. The 1D
and the 3D models predict different neutrino fluxes from the various
nuclear reactions, as shown in Table 1, where they are compared with
the experimental results obtained until now. As stated above, it is
not possible, at present, to discriminate the solutions due to model uncertainties and experimental errors. A
measurement of the CNO flux, with reasonable errors, could distinguish
between the two models which predict substantially different fluxes.
The pp solar neutrino flux has been never measured directly. Gallex
and Sage have measured the integrated solar flux from 233 keV, which,
together with the Borexino $^{7}$Be neutrino flux measurement and
the experimental data on the $^{8}$B neutrino flux, can be used to
infer the pp neutrino flux with a relatively small uncertainty, once
the luminosity constraint is applied (fig. \ref{Pee}). Nevertheless
a direct experimental observation, which can be compared with the
solar luminosity and the SSM prediction, would be an important achievement.
The pp flux measurement is part of the Borexino phase 2 program.

\subsection*{Improvement of the$^{7}$Be solar neutrino flux measurement}

Improving the $^{7}$Be flux measurement is one of the goals for Borexino
phase II. The physics goals of this study can be summarized in three
main points. 

\begin{enumerate}
\item 
Reduction of the total error (statistical+systematic) down to hopefully
$\sim$3\%. Even with such a precision this measurement
cannot solve unambiguously the metallicity puzzle, because of the uncertainties
of the SSM. The $^{7}$Be flux measured in phase I falls in between
the two predicted flux values, for high and low metallicities, and a smaller
uncertainty would not help if this will be the case also for phase
II. A very precise experimental determination of the $^{7}$Be flux
remains nonetheless an important tool for testing the Solar Model
as well as a remarkable technical achievement. 

\item 
In the context of the neutrino physics, the Non Standard neutrino
Interactions are currently debated. One way to study them is
to analyze the shape of the oscillation vacuum-matter transition region.
While $^{7}$Be cannot have a conclusive role in this matter, it can
nevertheless help in restricting the range of the NSI flavor diagonal
terms.

\item
It is possible to constraint the NSI parameters studying the $\nu-e$ 
elastic scattering. Bounds are imposed by various
other experiments on solar, atmospheric and reactor (anti)neutrinos.
But $^{7}$Be neutrinos have the strong advantage of being mono-energetic
($^{8}$B neutrino detected by the other solar experiments in real
time have a continuous energy spectrum). In Borexino, the limitation
to this analysis comes from the residual background, especially $^{85}$Kr,
and, to lesser extend, $^{210}$Bi, which can mimic non-zero values
of the NSI parameters $\epsilon_{\alpha L}$ and $\epsilon_{\alpha R}$. 
An increase of statistics does not help much if not accompanied
by a reduction of such background.

\end{enumerate}

\subsection*{pp Solar Neutrino measurement}

This is the most important target of opportunity for Phase 2. The
very low $^{85}$Kr and reasonably low $^{210}$Bi achieved, make
a direct pp measurement a reality. A careful understanding of the
spectrum response in the $^{14}$C end-point region is crucial, its
study is possible through a dedicated effort. The main problem in
the pp-neutrino study is the disentanglement of the very tail of the
$^{14}$C spectrum (with possible pile-up) from the pp- neutrino spectrum.
The feasibility of the measurement is under study. A direct detection of pp neutrinos would be
a spectacular result and would justify the phase II alone. 

The analysis group performed a study of sensitivity to pp solar neutrinos
with the current background levels achieved, the expected statistical precision
of the pp-neutrino flux measurement is below 10\%, the systematics error will be mainly connected with uncertainty of the
$^{14}C$ pile-up spectrum and intensity determination.

\subsection*{pep Solar Neutrino measurement}

The first indication for pep solar neutrinos has been reported by
the collaboration \cite{PEP11}. The value for the pep interaction
rate obtained in Phase I (590 live-days) was $3.1\pm0.6(stat)\pm0.3(syst)$
cpd/day/100 tones, the absence of a pep signal was rejected at 98\%
C.L. The current measurement, in conjunction with the SSM (the uncertainty
in the pep flux is as low as 1.2\%), yields a survival probability
of Pee = $0.62\pm0.17$ though the uncertainties are far from Gaussian.
The precision is dominated by the statistical uncertainty (about 20\%),
though with more (background-free) data, systematic uncertainties
(10\%) will start to become important.

This is an extremely important result, but shy of the first measurement
of pep solar neutrinos. The addition of a modest batch of data with
$^{210}$Bi reduced at or below 30 counts/(100 ton $\times$ day)
will result in the first measurement (3$\sigma$) of pep solar
neutrinos. A much prolonged data taking could also result in a 5$\sigma$
precision measurement. The measurement will allow to gauge the survival
probability in the immediate proximity of the transition between two
different oscillation regimes.

\subsection*{solar $^{8}$B neutrino flux measurements.}

The Borexino detector is the first large volume liquid scintillator
detector sensitive to the low-energy solar neutrinos. It possesses
a very good energy resolution in comparison to the water Cherenkov
detectors, what allows to search for the solar $^{8}$B neutrinos
starting practically from the energies of the so called Thallium limit
(maximum energy of $\gamma$ rays from the chains of radioactive decay
of $^{232}$Th and $^{238}$U; gamma-quantum with maximum energy E=2.6
MeV is emitted in the decay of $^{208}$Tl). The measurements of the
$^{8}$B above 2.8 MeV has been performed using one year statistics
(246 days of live- time) of the Borexino data \cite{B8}. The threshold
of 2.8 MeV is the lowest achieved so far in the $^{8}$B neutrino
real-time measurements. The interest in the neutrino flux measurement
with low threshold comes from the peculiar properties of the survival
probability in this energy region. The electron neutrino oscillations
at E<2 MeV are expected to be driven by the so called vacuum oscillation,
and at energies E>5 MeV - by resonant matter-enhanced mechanism. The
energy region in between has never been investigated in spectrometric
regime, and is of particular interest because of the expected smooth
transition between the two types of oscillations.

The rate of $^{8}$B solar neutrino interaction as measured through
their scattering on the target electrons is 0.22$\pm$0.04(stat)$\pm$0.01(sys)
cpd/100 tons. This corresponds to an equivalent electron neutrino
flux of $\Phi_{^{8}\text{B}}^{ES}$=(2.4$\pm$0.4$\pm$0.1)$\times$10$^{6}$
cm$^{-2}$s$^{-1}$, as derived from the elastic scattering only,
in good agreement with existing measurements and predictions. The
corresponding mean electron neutrino survival probability, assuming
the BS07(GS98) Standard Solar Model (High Z model), is 0.29$\pm$0.10
at the effective energy of 8.6 MeV. The ratio between the measured
survival probabilities for $^{7}$Be and $^{8}$B is 1.9$\sigma$
apart from unity (see Fig.\ref{Pee}). For the first time the presence
of a transition between the low energy vacuum-driven and the high-energy
matter-enhanced solar neutrino oscillations is confirmed using the
data from a single detector, the result is in agreement with the prediction
of the MSW-LMA solution for solar neutrinos. 

Acquiring more statistics (of up to 5 years of the calendar time)
the Borexino will provide the competitive measurement of the $^{8}$B
neutrino flux.

\subsection{SOX campaign}

The Borexino large size and possibility to reconstruct an interaction point (with a precision of 14~cm at 1 MeV energy deposit) makes it an appropriate tools for searching of sterile neutrinos. If the oscillation baseline is about 1 m (which corresponds to $\Delta m^{2}\sim$1 eV$^{2}$), exposure of the detector to a compact powerful neutrino source should give rise to a typical oscillation picture with dips and rises in the spatial distribution of events density with respect to the source. Right beneath the Borexino detector, there is a cubical pit (side 105 cm) accessible
through a small squared tunnel (side 95 cm) that was built at the time of construction with the purpose of housing neutrino sources for calibration purposes. Using this tunnel, the experiment with neutrino source can be done with no changes to the Borexino layout. The center of the pit is at 8.25 m from the detector center, requiring a
relatively high activity of the neutrino source in order to provide detectable effect. 

The experiment SOX (for Short distance neutrino Oscillations with BoreXino) \cite{SOX} will be carried in three stages with gradually increasing sensitivity:
\begin{description}
\item[Phase A] a $^{51}$Cr neutrino source of 200-400 PBq activity deployed at 8.25 m from the detector center (external with respect to the detector); 
\item[Phase B] deploying a $^{144}$Ce-$^{144}$Pr antineutrino source with 2-4 PBq activity at 7.15 m from the detector center (placed in the detector's water buffer);
\item[Phase C] a similar $^{144}$Ce-$^{144}$Pr source placed right in the center of the detector. 
\end{description}

\begin{figure}[t]
\begin{center}
\includegraphics[width=0.80\textwidth]{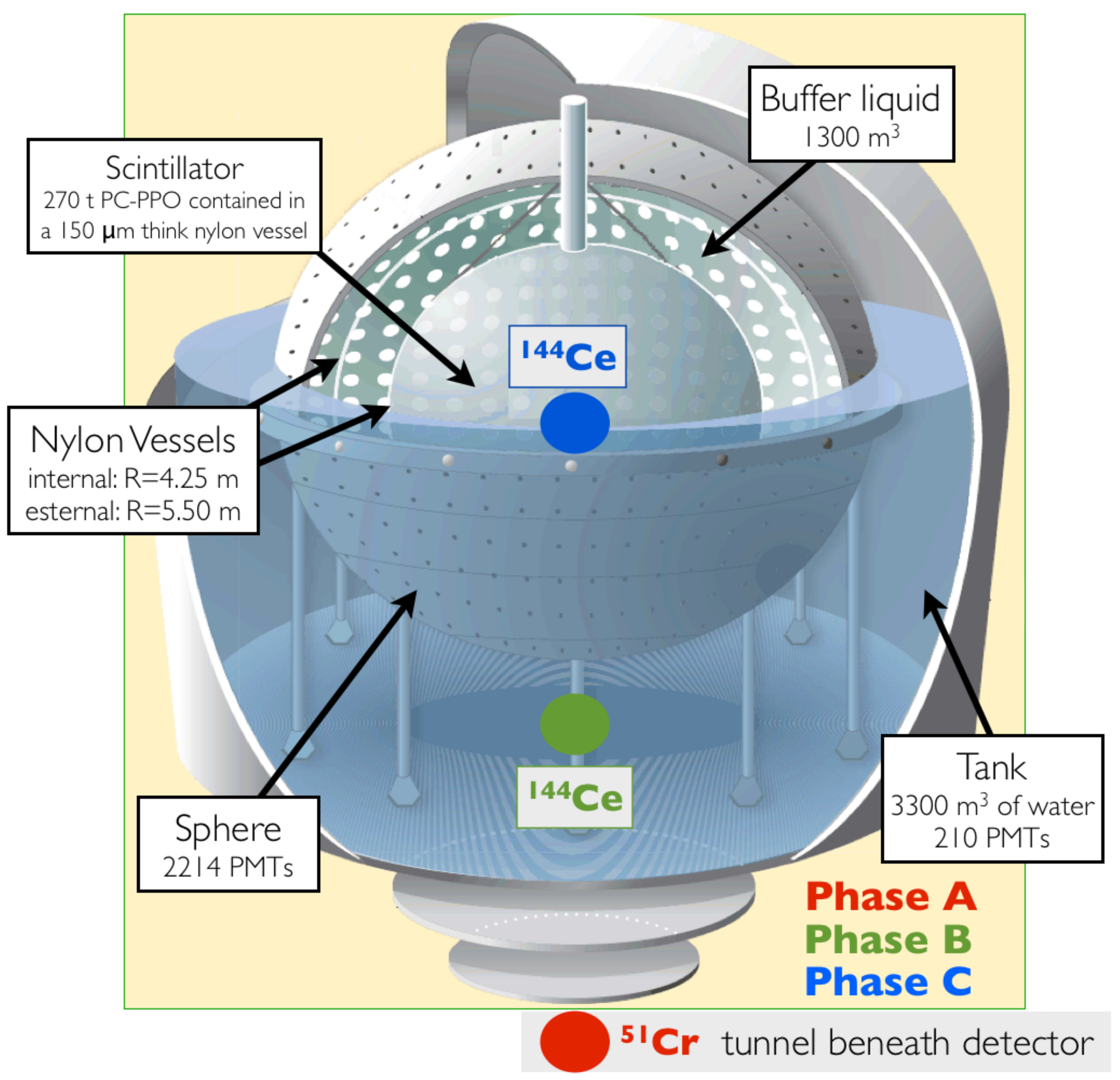}
\caption{\label{fig:detector} Layout of the Borexino detector and the approximate location of the neutrino and anti-neutrino sources in the three phases.}
\end{center}
\end{figure}

Figure 1 shows a schematic layout of the Borexino detector and the approximate location of the neutrino and anti-neutrino sources in the three phases. Two types of neutrino sources are considered: the $^{51}$Cr source and the $^{144}$Pr based source. $^{51}$Cr decays via electron capture into $^{51}$V, emitting two neutrino lines of 750 keV (90\%) and 430 keV (10\%), while $^{144}$Pr decays $\beta$ into $^{144}$Nd with an end--point of 3 MeV (parent $^{144}$Ce decays too, but end-point of its $\beta$-decay is below the IBD threshold). The portion of the $^{144}$Pr spectrum above the 1.8 MeV detection threshold is the only of importance for the experiment. Elastic scattering of $\bar{\nu}_e$ on electrons induce negligible background.

The source activity of 200-400 PBq is challenging, but only a factor 2-4 higher than what already done by Gallex and SAGE in the 90's. The $^{144}$Ce--$^{144}$Pr experiment in Phases B and C doesn't require high source activity. 
The Phase C is the most sensitive but it can  be done only after the shutdown of the solar neutrino program, because it needs modification of the detector. The Phases A and B will not disturb the solar neutrino program of the experiment, which is supposed to continue until the end of 2015, and do not require any change to Borexino hardware. The challenge for the Phase C is constituted by the large background induced by the source in direct contact with the scintillator, that can be in principle tackled thanks to the correlated nature of the $\bar{\nu}_e$ signal detection. In Phase B this background, though still present, is mitigated by the shielding of the buffer liquid.

Borexino can study short distance neutrino oscillations in two ways: by comparing the detected number of events with expected value (disappearance technique, or total counts method), or by observing the oscillation pattern in the events density over the detector volume (waves method). In the last case the typical oscillations length is of the order The variations in the survival probability $P_{ee}$ could be seen on the spatial distribution of the detected events as the waves superimposed on the uniformly distributed background. Oscillation parameters can be directly extracted from the analysis of the waves. The result may be obtained only if the size of the source is small compared to the oscillation length. The $^{51}$Cr source will be made by about 10-35 kg of highly enriched Cr metal chips which have a total volume of about 4-10 l. The source linear size will be about 15-23 cm, comparable to the spatial resolution of the detector. The $^{144}$Ce--$^{144}$Pr source is even more compact. All simulations shown below takes into account the source size.

In Phase A the total counts method sensitivity is enhanced by exploiting the fact that the life-time of the $^{51}$Cr is relatively short. In Phases B and C this time-dependent method is not effective because the source life-time is longer (411 days), but this is compensated by the very low background and by the larger cross-section. The total counts and waves methods combined together yield a very good sensitivity for both experiments. Besides, the wave method is independent on the intensity of the source, on detector efficiency, and is potentially a nice probe for un-expected new physics in the short distance behavior of neutrinos or anti-neutrinos. 

\begin{figure}[t]
\begin{center}
\includegraphics[width=0.95\textwidth]{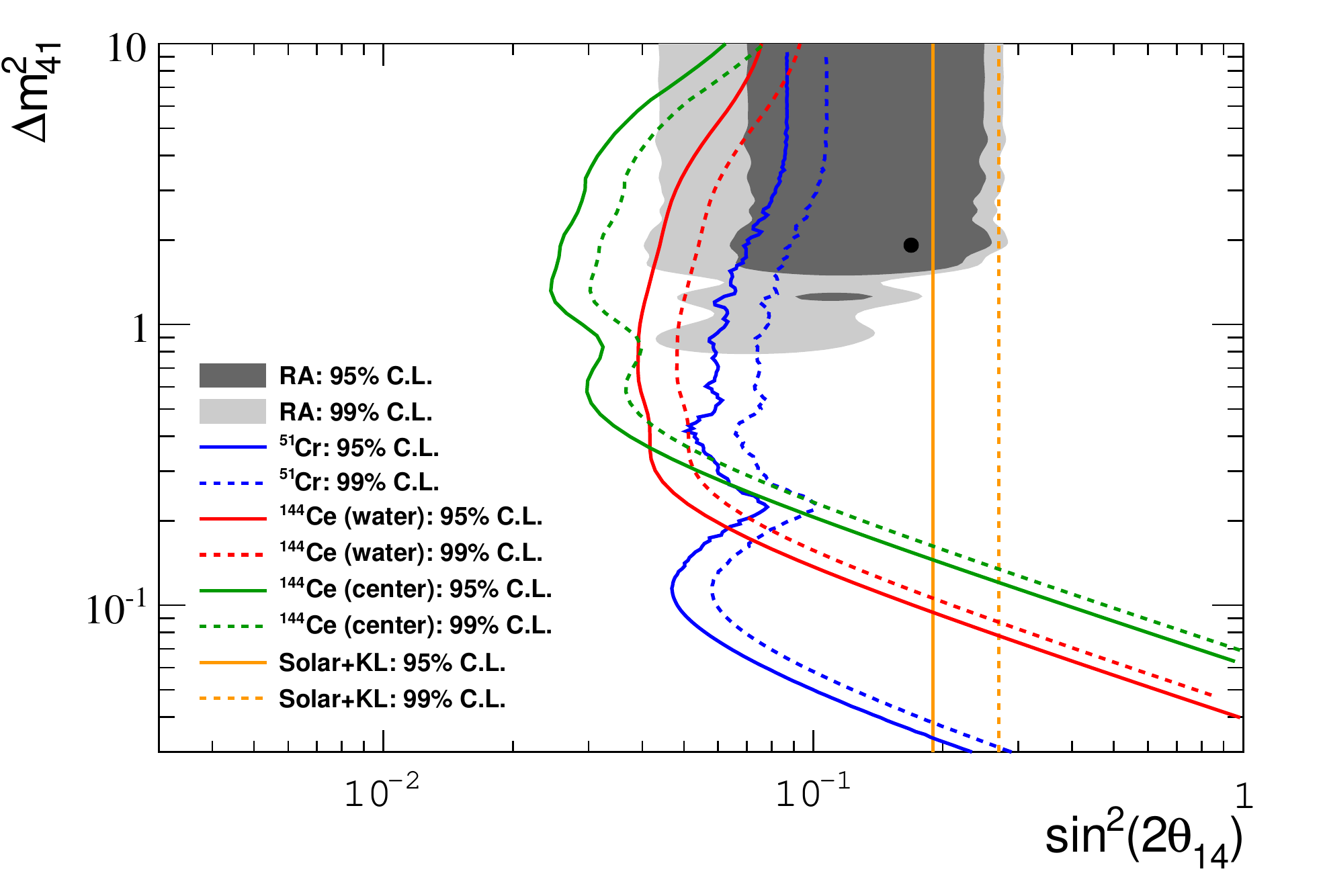}
\end{center}
\caption{\label{fig:sensitivity} Sensitivity of the Phase A ($^{51}$Cr external, blue), of Phase B ($^{144}$Ce--$^{144}$Pr external, red) and Phase C ($^{144}$Ce--$^{144}$Pr center, green). The gray area is the one indicated by the reactor anomaly, if interpreted as oscillations to sterile neutrinos. Both 95\% and 99\% C.L. are shown for all cases. The yellow line indicates the region already excluded in \cite{bib:palazzo2}.}
\end{figure}

The sensitivity of SOX with respect to oscillation into sterile neutrino was evaluated with a toy Monte Carlo. Expected statistical samples (2000 events) were generated for each pair of oscillation parameters.  We assume a period of 15 weeks of stable data taking before the source insertion in order to accurately constrain the background. The background model includes all known components, identified and accurately measured during the first phase of Borexino. We built the confidence intervals from the mean $\chi^2$ for each couple of parameters with respect to the non-oscillation scenario. The result is shown in Fig.~\ref{fig:sensitivity}, as one can see fro the figure the reactor anomaly region of interest is mostly covered. 

The simulation for the Phase A are shown for a single irradiation of the $^{51}$Cr source up to the initial intensity of 370 PBq (10 MCi) at the site. A similar result can be obtained with two irradiations of about 200 PBq if higher intensity turns out to be beyond the technical possibilities. The single irradiation option is preferable and yields a slightly better signal to noise ratio.

The physics reach  for the $^{144}$Ce--$^{144}$Pr external (Phase B) and internal (Phase C) experiments, assuming 2.3 PBq (75 kCi) source strength and one and a half year of data taking) is shown in the same figure (Fig.~\ref{fig:sensitivity}).  

The $\chi^2$ based sensitivity plots are computed assuming significantly bigger volume of liquid scintillator (spherical vessel of 5.5 m radius), compared to the actual volume of liquid scintillator (limited by a sphere with 4.25 m radius) used for the solar phase. Such an increase will be made possible by the addition of the scintillating fluor (PPO) in the inner buffer region (presently inert) of the detector.
We have also conservatively considered exclusion of the innermost sphere of 1.5 m radius from the analysis in order to reject the gamma and bremsstrahlung backgrounds from the source assembly. Under all these realistic assumptions, it can be noted from Fig.~\ref{fig:sensitivity} that the intrinsic $^{144}$Ce--$^{144}$Pr sensitivity is very good: for example the 95\% C.L. exclusion plot predicted for the external test covers adequately the corresponding reactor anomaly zone, thus ensuring a very conclusive experimental result even without deploying the source in the central core of the detector.

\section{Conclusions}

Borexino achieved its main goal, but is still a challenging detector. Due to achieved level of purification (much higher than it was needed for $^{7}$Be neutrino detection) some measurements of solar neutrino fluxes beyond the original program were performed. Borexino-II is operating with repurified LS, at new levels of radiopurity, two years of data are collected and are being analyzed aiming the pp-neutrino flux and CNO neutrino flux measurement. Borexino is also an ideal detector to test the sterile neutrinos through the disappearance/wave effects. The proposed staged approach ($^{51}$Cr source at the first stage and  two $^{144}$Ce–$^{144}$Pr experiments) is a comprehensive sterile neutrino search which will either confirm the effect or reject it in a clear and unambiguous way. 
In particular, in case of one sterile neutrino with parameters corresponding to the central value of the reactor anomaly, SOX will surely discover the effect, prove the existence of oscillations and measure the parameters through the “oscillometry” analysis.

\section*{Acknowledgments}
Borexino was made possible by funding from INFN (Italy), NSF (USA), BMBF, DFG, and MPG (Germany), NRC Kurchatov Institute (Russia), MNiSW (Poland, Polish National Science Center (grant DEC-2012/06/M/ST2/00426)), Russian Foundation for Basic Research (Grant 13-02-92440 ASPERA, the NSFC-RFBR joint research program) and RSCF research program (Russia). We acknowledge the generous support of the Gran Sasso National Laboratories (LNGS). SOX is funded by the European Research Council.

\end{document}